\documentclass[usenatbib]{mnras}
\usepackage{CJKutf8}
\usepackage{newtxtext,newtxmath}

\usepackage[T1]{fontenc}
\usepackage{verbatim}
\usepackage{color}
\usepackage[switch]{lineno}
\usepackage{subcaption}
\usepackage{float}
\usepackage{ulem}

\DeclareRobustCommand{\VAN}[3]{#2}
\let\VANthebibliography\thebibliography
\def\thebibliography{\DeclareRobustCommand{\VAN}[3]{##3}\VANthebibliography}

\usepackage{graphicx}	
\usepackage{amsmath}	
\usepackage{siunitx}
\usepackage{threeparttable}

\newcommand{\gsim}{\mbox{\hspace{.2em}\raisebox{.5ex}{$>$}\hspace{-.7em}\raisebox{-.5ex}{$\sim$}\hspace{.2em}}}

\newcommand{\ssst}{\scriptscriptstyle}
\newcommand{\E}[1]{\times 10^{#1}}

      \newcommand{\ps}{\,{\rm s}^{-1}}
\newcommand{\yr}{\,{\rm yr}}    \newcommand{\Msun}{M_{\odot}}
\newcommand{\cm}{\,{\rm cm}}    \newcommand{\km}{\,{\rm km}}

\newcommand{\parsec}{\,{\rm pc}}\newcommand{\kpc}{\,{\rm kpc}} 
\newcommand{\erg}{\,{\rm erg}}        \newcommand{\K}{\,{\rm K}}
    \newcommand{\keV}{\,{\rm keV}}

\newcommand{\Ts}{T_{s}}
\newcommand{\rs}{r_{s}}         \newcommand{\vs}{v_{s}}
        \newcommand{\mH}{m_{\ssst\rm H}}
\newcommand{\nHH}{n({\rm H}_{2})} 
\newcommand{\VLSR}{V_{\ssst\rm LSR}}
\newcommand{\xray}{X-ray}       
\newcommand{\gray}{$\gamma$-ray}

\newcommand{\snr}{Kes\,67}
\newcommand{\twCO}{$^{12}$CO}   \newcommand{\thCO}{$^{13}$CO}
\newcommand{\HCOp}{HCO$^+$}
\newcommand{\Jotz}{$J$=1--0}    

\title[Molecular Shell and GeV emission of SNR Kes\,67]{
  Expanding Molecular Shell and Possible \gray\ Source Associated with Supernova Remnant Kesteven\,67
 }

\author[Y.Z.Shen et al.
]{
Yun-Zhi Shen 
(\begin{CJK}{UTF8}{bsmi}沈蘊之\end{CJK})
$^{1}$,
Yang Chen 
(\begin{CJK}{UTF8}{bsmi}陳陽\end{CJK})
$^{1,2}$, \thanks{E-mail: ygchen@nju.edu.cn}
Xiao Zhang 
(\begin{CJK}{UTF8}{bsmi}張瀟\end{CJK})
$^{3,1,2}$, \thanks{E-mail: xiaozhang@njnu.edu.cn}
Tian-Yu Tu 
(\begin{CJK}{UTF8}{bsmi}涂天宇\end{CJK})
$^{1}$, 
\newauthor{
Wen-Juan Zhong 
(\begin{CJK}{UTF8}{bsmi}鍾文娟\end{CJK})
$^{1}$,
Qian-Qian Zhang 
(\begin{CJK}{UTF8}{bsmi}張芊千\end{CJK})
$^{1}$,
and 
Qian-Cheng Liu 
(\begin{CJK}{UTF8}{bsmi}劉前程\end{CJK})
$^{1}$}{}
\\
$^{1}$School of Astronomy \& Space Science, Nanjing University, Nanjing 210023, China\\
$^{2}$Key Laboratory of Modern Astronomy and Astrophysics, Nanjing University, Ministry of Education, Nanjing 210023, China\\
$^{3}$School of Physics and Technology, Nanjing Normal University, Nanjing 210023, China
}

\date{Accepted XXX. Received YYY; in original form ZZZ}

\pubyear{0000}

\begin{document}
\label{firstpage}
\pagerange{\pageref{firstpage}--\pageref{lastpage}}
\maketitle

\begin{abstract}
We investigate the molecular environment of the supernova remnant\,\,(SNR) Kesteven\,67 (G18.8+0.3) using observations in  \twCO,\,\thCO,\,\HCOp,\,and\,HCN lines and possible associated \gray\ emission using 16-yr {\sl Fermi}-LAT observation.
We find that the SNR is closely surrounded by a molecular belt in the southeastern boundary, with the both recessed in the band-like molecular gas structure along the Galactic plane.
The asymmetric molecular line profiles are widely present in the surrounding gas around local-standard-of-rest velocity $+20\km\ps$. 
The secondary components centered at $\sim+16\km\ps$ in the belt and $\sim+26\km\ps$ in the northern clump can be ascribed to the motion of a wind-blown molecular shell.
This explanation is supported by the position-velocity diagram along a line cutting across the remnant, which shows an arc-like pattern, suggesting an expanding gas structure.
With the simulation of chemical effects of shock propagation, the abundance ratios $N$(\HCOp)/$N$(\twCO) $\sim2.6\E{-5}$--$3.6\E{-4}$ obtained in the belt can be more naturally interpreted by the wind-driven bubble shock than by the SNR shock.
The belt and northern clump are very likely to be parts of an incomplete molecular shell of bubble driven by O-type progenitor star's wind.
The analysis of 0.2--500\,GeV \gray\,emission uncovers a possible point source (`Source~A') about 6.5$\sigma$ located in the north of the SNR, which essentially corresponds to northern molecular clump. 
Our spectral fit of the emission indicates that a hadronic origin is favored by the measured Galactic number ratio between CR electrons and protons $\sim0.01$.
\end{abstract}




\section{Introduction}
Investigation of molecular environment of supernova remnants (SNRs) is of great importance in understanding the dynamical evolution of the SNR, hadronic \gray\ emission \citep{hadronic_pion_decay}, ionizing low energy cosmic ray (CR) protons \citep{CR280MeV}, the pre-supernova (SN) stellar wind \citep{wind_bubble_2013}, effects of shock chemistry \citep{shock_chemistry}, as well as the nearby triggered star formation \citep[e.g.][]{trigger_star_formation}.

In an SNR-molecular cloud (MC) association, it is very likely that a molecular gas bubble is blown inside the MC by the  progenitor system of either core collapse or even Type Ia SN before the SN blastwave hits the MC or the molecular bubble shell \citep[see e.g., ][]{wind_bubble_2013, Tycho_shell, N103B}.
Therefore, the role of the wind-blown molecular bubble is unavoidable in studying the physical and chemical properties of the SNR.
In this regard, Kes\,67 is a typical object, in which the interplays of the progenitor's bubble shell with both the ambient medium and SNR shock deserves deep examining.

SNR Kesteven\,67 (G18.8+0.3; hereafter \snr\ for short) has an unusual radio morphology, which is bright in the east and south and fades westward. 
At the southern edge, the radio emission displays a nearly right angle. 
The local-standard-of-rest (LSR) velocity center of this complex is at +19$\km\ps$ \citep{Dubner1999, paron_south_clump}
.
The previous study of molecular environment of the remnant showed a positional agreement between the SNR shock and eastern elongated molecular feature. The far-infrared radiating dust was also detected around the SNR correlating with the molecular feature and supporting the shock-heated origin \citep{Dubner1999}. 
\citet{paron_south_clump} pointed out a coincidence between the indentations of the SNR radio continuum emission and the protrusions in the MC at the remnant's southern boundary.
These results strongly imply that the SNR shock have already interacted with MCs.

In addition to the MCs along the boundary of SNR, there are farther clumps to the east and south of the SNR, respectively. The eastern clump is suggested to be shaped by the HII region inside and have not been contacted by the shock front \citep{Paron_kes67_clump}.
The further southern molecular clumps, which are not corresponding to the position of \twCO\ peak, partially surround an HII region \citep{paron_south_clump}.

The Fe I K$\alpha$ line of 6.3-6.5\,keV has been detected in \snr\ \citep{Fe}, indicating the penetration of low-energy  CRs (LECRs) into the dense gas.
This SNR is a soft X-ray source with an electron temperature of $\sim$0.4\,keV \citep{Fe}.

By means of CO emission and HI absorption features, the distance of Kes\,67 has been constrained at the far-end from Earth and located at 13.8$\pm$0.4\kpc\ \citep{age, HI_distance,distance_new}.

In this work, we study the molecular environment of \snr\ using emission lines of molecular species \twCO, \thCO, \HCOp, and HCN and search for \gray\ emission associated with the SNR. 
We present the spatial distribution and line profiles of the eastern molecular gas, calculate the column density ratio between \HCOp\ and CO, and use the Paris-Durham shock code to discuss the shock condition for the obtained abundance ratio.
We analyse the 16 years of {\sl Fermi}-LAT observation data and find \gray\ emission in energy range 0.2--500\,GeV in the north of the remnant. The observation data are described in Section~\ref{sec:observation} and the analysis results are given in Section~\ref{sec:results}. The results are discussed in Section~\ref{sec:discussion} and conclusions are in Section~\ref{sec:conclusion}.

\section{Observations and data}\label{sec:observation}

\subsection{Millimeter Wave Molecular Line Data}

We use the 13.7\,m millimeter-wavelength telescope of the Purple Mountain Observatory at Delingha (hereafter PMOD), China, in 2020 June, to perform observation towards SNR Kes\,67 in the \HCOp\,(\Jotz) line at \SI{89.188}{GHz} and the HCN\,(\Jotz) line at \SI{88.632}{GHz}. The observation uses the on-the-fly (OTF) mapping mode. The half-power beam width (HPBW) is $\approx60''$. We map a 22$^{\prime}$ $\times$22 $^{\prime}$ area centred at (18$^{\text{h}}$23$^{\text{m}}$53$^{\text{s}}$.83, $-
$12$^{\circ}$30$'$11$''$.70, J2000.0), which includes most of the southern side of the remnant via raster-scan mapping with a grid spacing of 30$^{\prime \prime}$. 
We make the beam correction with the main-beam efficiency of 0.628. Using the GILDAS/CLASS package\footnote{\url{http://www.iram.fr/IRAMFR/GILDAS}}, the velocity resolution of both spectra was 0.25\,km\,s$^{-1}$ and the $\VLSR$ range was $-$100\,km\,s$^{-1}$ to +150\,km\,s$^{-1}$, and the pixel size was 30$''$. The RMS noise is $\sim$ 0.1 K for \HCOp\ and HCN.

We also use the archival data of the \twCO\ \,(\Jotz) line at \SI{115.271}{GHz} and the \thCO\ \,(\Jotz) line at \SI{110.201}{GHz} of the FOREST Unbiased Galactic plane Imaging survey with the Nobeyama 45 m telescope (FUGIN; \citet{2017PASJ...69...78U}) observation. 
The angular resolution was 20$^{\prime\prime}$ for \twCO\ and 21$^{\prime\prime}$ for \thCO\ . The average rms noise was $\sim$\SI{1.5}{\kelvin} for \twCO\ and $\sim$\SI{0.7}{\kelvin} for \thCO\ at a velocity resolution of 0.65\,km\,s$^{-1}$. 

\subsection{{\sl Fermi}-LAT \gray\ Data}
For the $\gamma$-ray emission, we use about 16 yr observation data of Large Area Telescope\,(LAT) onboard the {\sl Fermi Gamma-ray Space Telescope}. The time frame of our research is from 2008-08-04 15:43:36 (UTC) to 2024-07-21 02:17:41 (UTC), and the circular region of interest (ROI) is 15$^{\circ}$ in radius, centred at the coordinates R.A.=275.96$^{\circ}$, Dec=$-$12.4$^{\circ}$ (J2000).
We use the standard software {\small Fermipy}\footnote{\url{https://fermipy.readthedocs.io/en/stable/}} (Vertion 1.2.0 released on 2022 September 21), which bases on the {\small Fermitools} \footnote{\url{http://fermi.gsfc.nasa.gov/ssc/data/analysis/software/}}(Version 2.2.0 realesed on 2022 June 21) to analyze the data.
We select `SOURCE' class (evclass=128, evtype=3) with the instrument response function (IRF) `P8R3\_SOURCE\_V3\_v1' and constrain the energy range to 0.2-500\,GeV. To eliminate the Earth's limb, we limit the maximum of zenith to 90$^{\circ}$. 
We also apply the recommended filter string `(DATA\_QUAL>0)\&\&(LAT\_CONFIG==1)' to choose the good time intervals. 
To build the background model, we use the {\sl Fermi}-LAT Fourth Source Catalog Data Release 4\,(\citet{Abdollahi_2020}, 4FGL-DR3 \citet{2022ApJS..260...53A}, 4FGL-DR4 \citet{4FGL-DR4}) in a radius of 15$^{\circ}$ around the center of the ROI to consider the \gray\ sources, as well as the Galactic diffuse emission (\textit{gll\_iem\_v07.fits}) and the isotropic emission (\textit{iso\_P8R3\_SOURCE\_V3\_v1.txt}).

\subsection{Other Data}
We use the Multi-Array Galactic Plane Imaging Survey (MAGPIS) continuum image with angular resolution of 6$''$ at 1.4GHz \citep{MAGPIS} to delineate the SNR boundary.

\section{Data Analysis and Results}\label{sec:results}
\subsection{Millimeter-wavelength Molecular Lines}\label{sec:molecular lines}
\subsubsection{Molecular Gas Distribution}\label{sec:molecular distribution}

\begin{figure*}
	\includegraphics[width=0.96\textwidth]{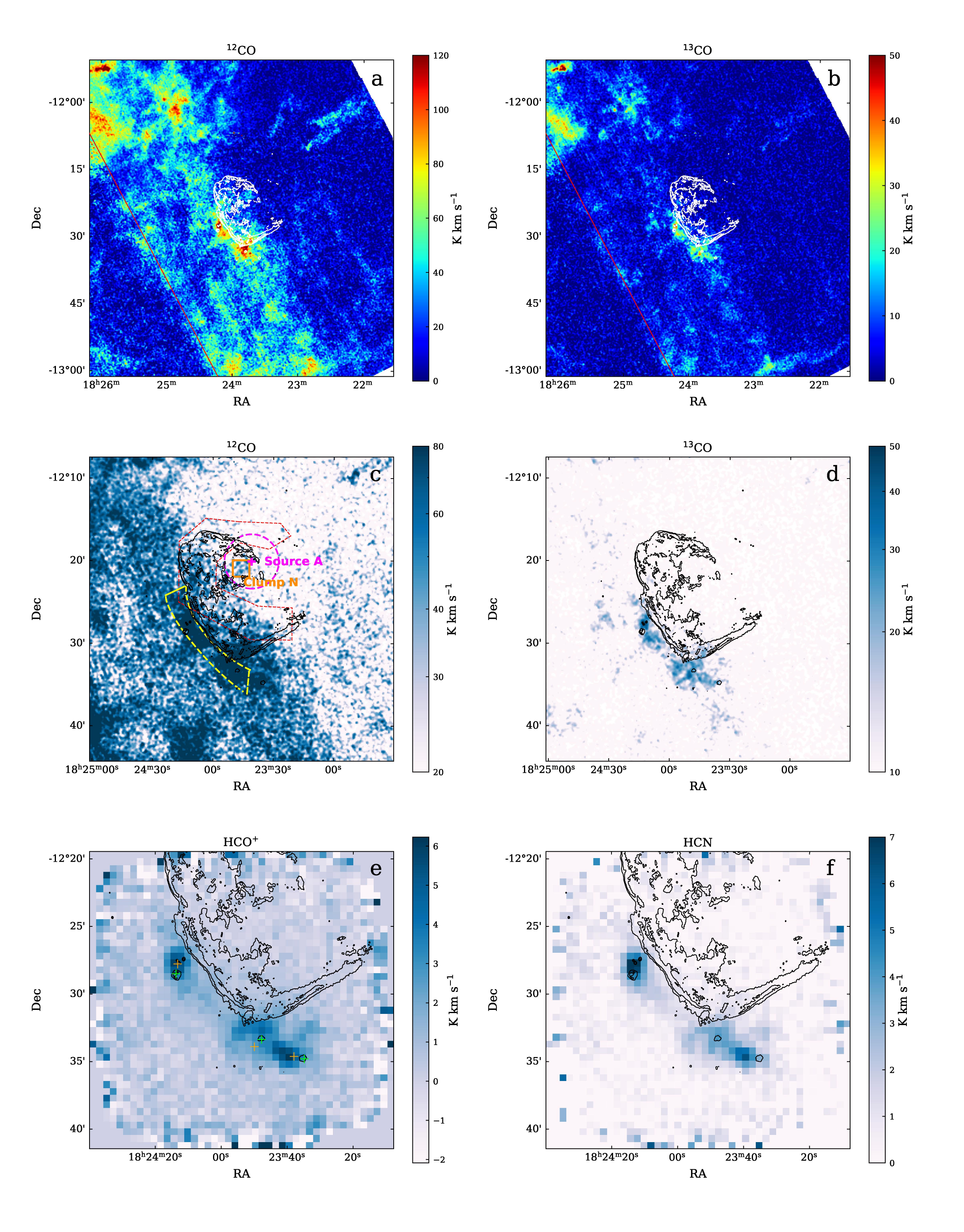}
    \caption{ 
    {\sl a, b}): The integrated-intensity pseudo-color image of \twCO\ \,(\Jotz) (left) and \thCO\ \,(\Jotz) (right) in the $\VLSR$ range from +10\,km\,s$^{-1}$ to +30\,km\,s$^{-1}$ in a large field of view, overlaid by the contours of 1.4\,GHz radio continuum of 2, and 5 mJy beam$^{-1}$. The same contours are also plotted in other four panels {\sl c, d, e, f)} in black. The solid red line masks the Galactic plane ($b=0$). 
    {\sl c, d}): The same as the top row, but in a small field of view and in greyscale. The orange box marks the position of Clump N. The magenta cross and dashed circle mark the position and 68\% positional uncertainty of \gray\ source `Source A' (see \S\ref{sec:fermi}), and the dashed red polygon is the region of the enhanced Fe\,I K$\alpha$ line emission of Kes\,67 adapted from Figure\,1(b) in \citet{Fe} .The dashed yellow lines surround the belt. 
    {\sl e, f}): Integrated-intensity maps of \HCOp\ and HCN from the PMOD observation in the $\VLSR$ range in +10 -- +30\,km\,s$^{-1}$. 
    The orange and green crosses represent ATLASGAL sources and HII regions, respectively.
    \label{fig: Integerated maps}
    }
\end{figure*}

\begin{figure*}
	\includegraphics[scale=.4]{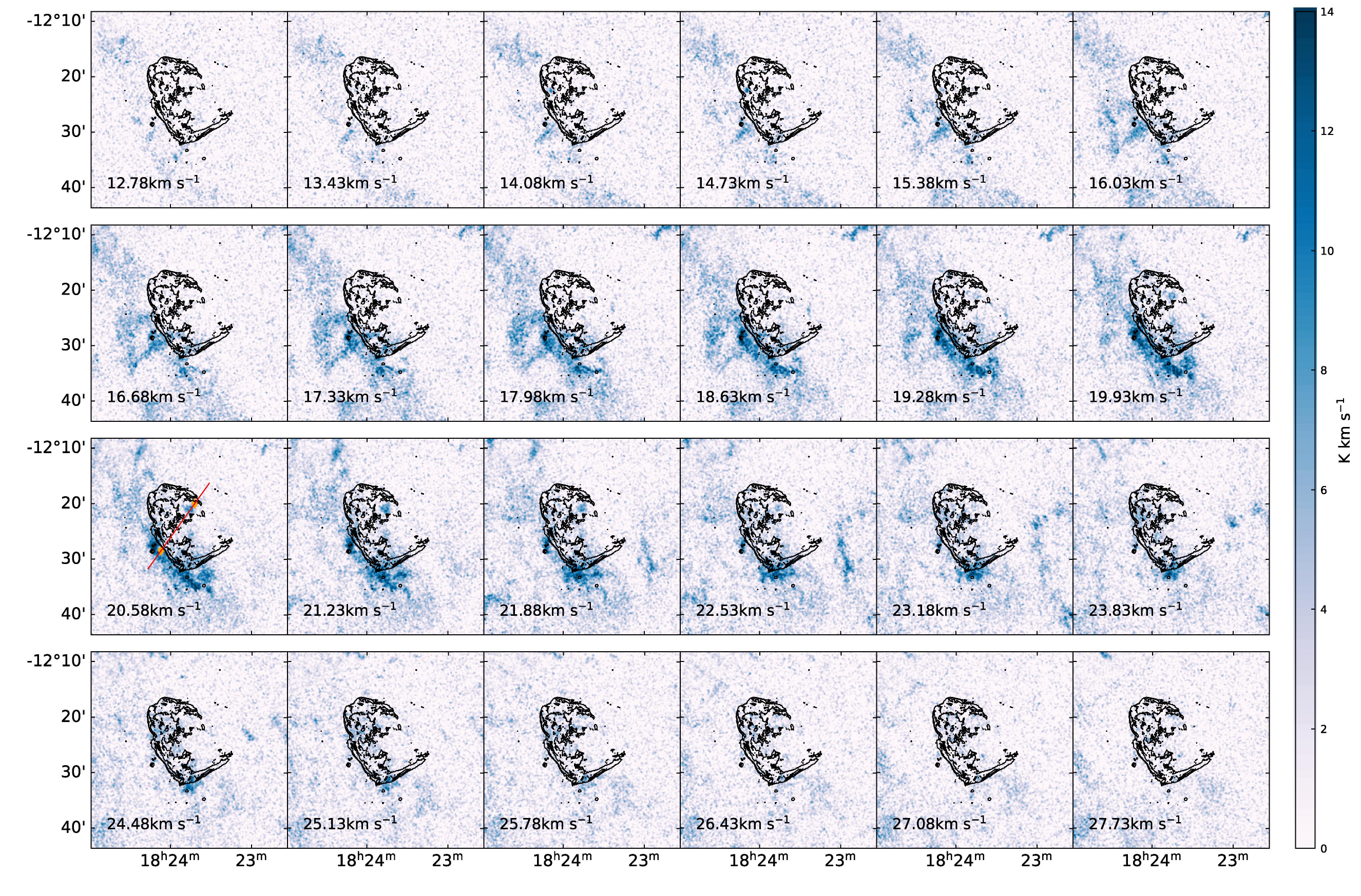}
    \caption{Velocity channel map of FUGIN \twCO\ \,(\Jotz) emission on a scale of +12.8\,km\,s$^{-1}$ to +27.7\,km\,s$^{-1}$ with a step of  0.65\,km\,s$^{-1}$, along with the contours of VLA 1.4GHz radio emission. 
    }
    \label{fig:velocity_channel_12CO} 
\end{figure*}

\begin{figure*}
	\includegraphics[width=0.96\textwidth]{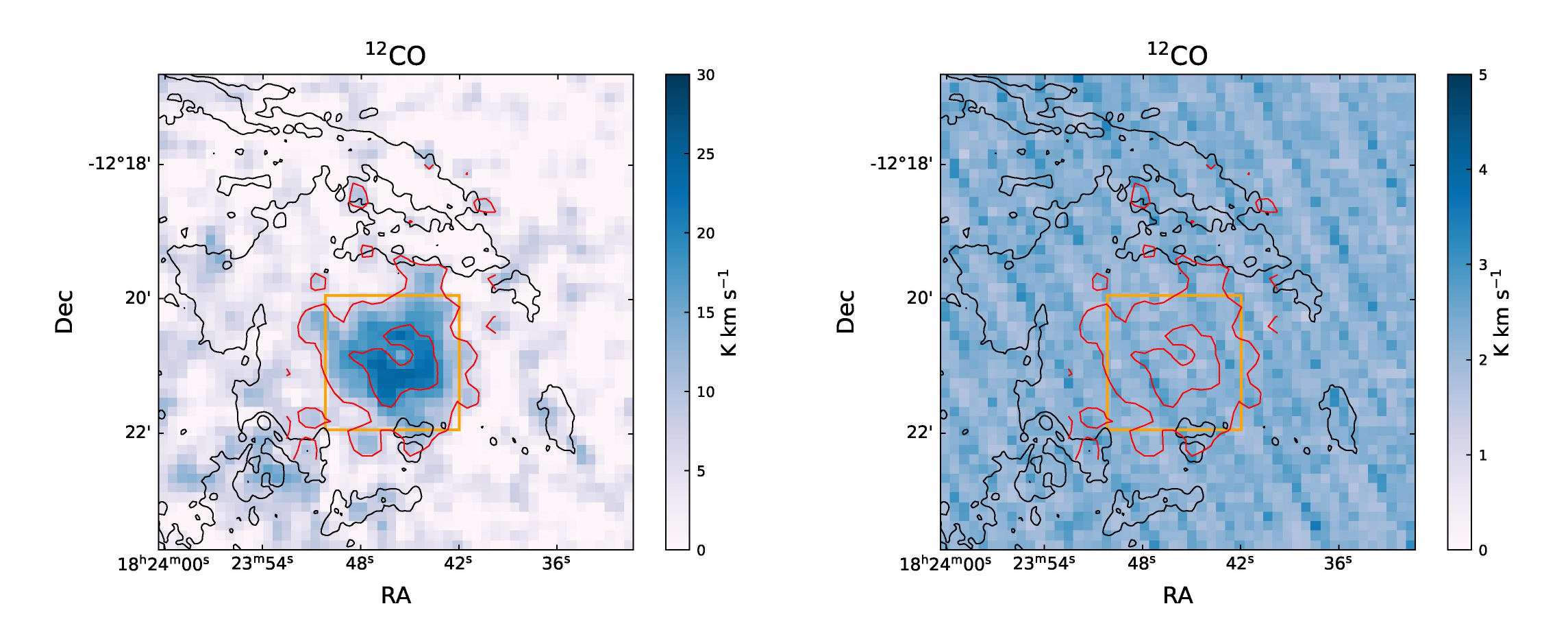}
    \caption{Integrated-intensity greyscale image of \twCO\ (\Jotz) of Clump N in the $\VLSR$ range +20 -- +22\,km\,s$^{-1}$  (left panel) and 1 $\sigma$ level noise of integrated map (right panel), overlaid with the 1.4\,GHz radio continuum contours of the SNR (in black) and the \twCO\ (\Jotz) contours  of Clump N (in red). 
    The orange box (the same as that in Figure~\ref{fig: Integerated maps}c) marks the spatial range for extracting CO line spectra of Clump N.}
    \label{fig: fig1_cont}

\end{figure*}

The top row of Figure~\ref{fig: Integerated maps} shows that SNR \snr\ is projected at the northwestern edge of a large band-like molecular gas structure, $\sim25'$ in thickness, above the Galactic plane ($b=0^{\circ}$), which is evident in the CO lines in the $\VLSR$ range from +10\,km\,s$^{-1}$to +30\,km\,s$^{-1}$.

In the close-up CO maps (Figures~\ref{fig: Integerated maps}c and \ref{fig: Integerated maps}d), the line emissions are bright in an outer belt closely along the southeastern edge of the remnant. 
There are two concentrations in the `belt', one in the east and the other in the south. 
The two concentrations 
are also seen in the \HCOp\ and HCN maps with the same $\VLSR$ range
(Figures~\ref{fig: Integerated maps}e and \ref{fig: Integerated maps}f). 
However, the brightness peaks of the eastern concentration is about $2'$ from the eastern radio boundary.
The eastern brightness peak appears to correspond to the eastern infrared clump that was discovered and suggested to contain proto-stars by \citet{Paron_kes67_clump}.
The southern peak can be divided into two clumps, which are related to the HII regions G018.630+0.309 (R.A. = $18^\text{h}23^\text{m}47.8^\text{s}$, Dec = $-12^\circ33'21''$) and G018.584+0.334 (R.A. = $18^\text{h}23^\text{m}34.9^\text{s}$, Dec = $-12^\circ34'50''$) 
\citep{HII_regions_catalog}\ and 
also related to ATLASGAL sources AGAL018.626+00.297 (R.A. = $18^\text{h}23^\text{m}49.8^\text{s}$, Dec = $-12^\circ33'53.8''$) and AGAL018.593+00.334 (R.A. = $18^\text{h}23^\text{m}38.0^\text{s}$, Dec = $-12^\circ34'38.0''$) \citep{ATLASGAL}, respectively. The former HII region and ATLATGAL source have been noticed in \citet{paron_south_clump}.

In Figure~\ref{fig:velocity_channel_12CO} we present the channel maps of \twCO\ \,(\Jotz)  in the velocity range from +12.78\,km\,s$^{-1}$ to +27.73\,km\,s$^{-1}$. 
In the range of +18\,km\,s$^{-1}$ -- +21\,km\,s$^{-1}$, the molecular gas in the east and south in the field of view including the `belt' has a sharp interface along the southeastern front of SNR shock wave represented by the radio contours. From +22\,km\,s$^{-1}$ to +24\,km\,s$^{-1}$, the southern molecular concentration appears to be coincident with the right-angle vertex of the radio boundary.
A molecular clump (denoted as `Clump N') can be discerned in the north of the SNR in the velocity range +20\,km\,s$^{-1}$ -- +22\,km\,s$^{-1}$, which is also obvious in the \twCO\ integrated map (Figure~\ref{fig: Integerated maps}c) and seems to be extended from the `belt'. 
Figure~\ref{fig: fig1_cont} shows the close-up towards Clump N presenting in \twCO\ (\Jotz) line, which is significant above 1 $\sigma$ level noise.

\subsubsection{Molecular Line Profiles}\label{sec:lines}

\begin{figure*}
	\includegraphics[width=1.0\textwidth,height=1.0\textwidth]{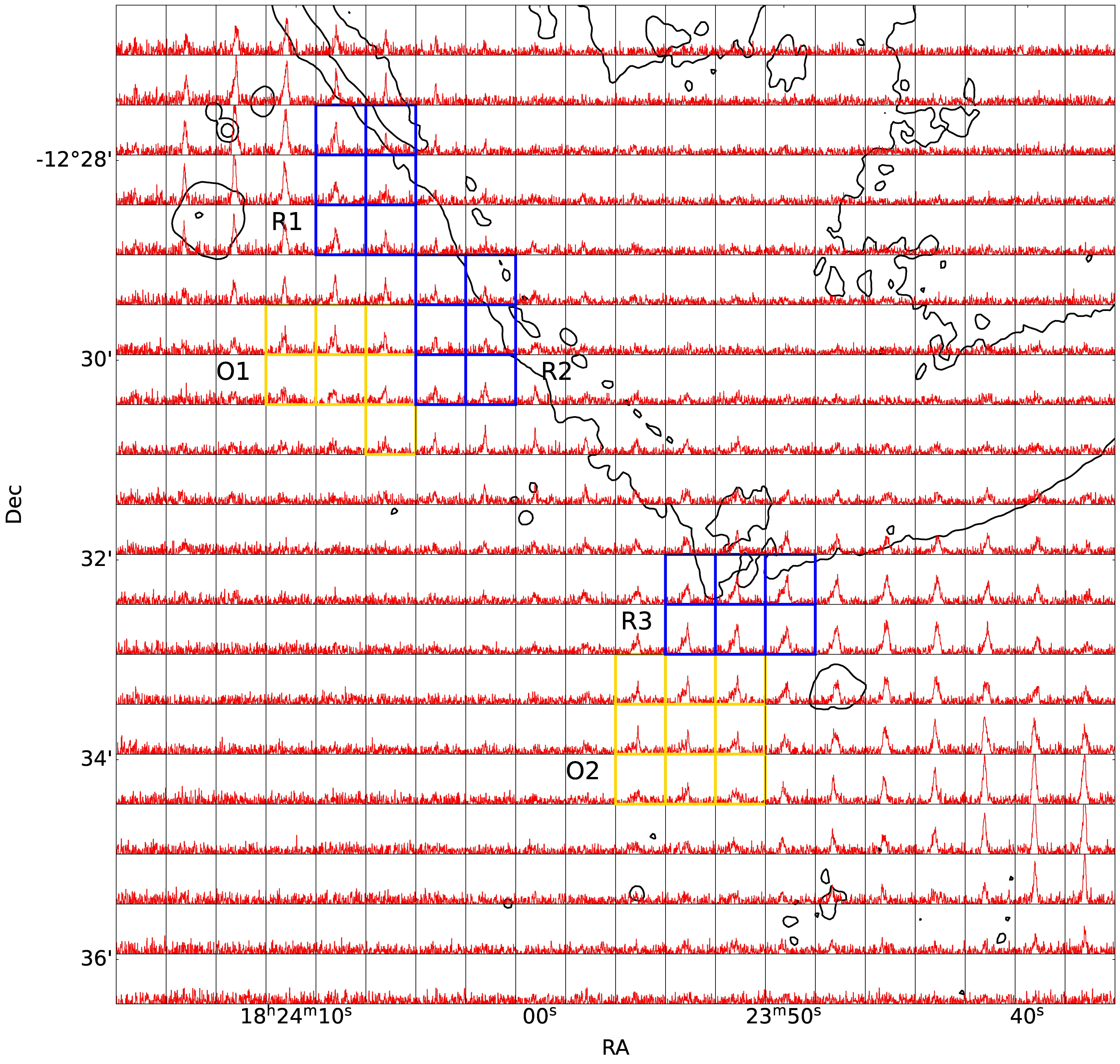}
    \caption{Grid map of \HCOp\ line profiles in the LSR velocity scale of 0\,km\,s$^{-1}$ -- +50\,km\,s$^{-1}$ with the black contours of VLA 1.4GHz radio emission. The size of each pixel is 30$^{\prime\prime}$$\times$30$^{\prime\prime}$. Regions R1, R2 and R3 are marked with blue rectangles and regions O1 and O2 are marked with yellow rectangles, from which molecular line the spectra are extracted.}
   \label{fig:gridmap}
\end{figure*}

\begin{figure*}
	\includegraphics[width=0.9\textwidth]{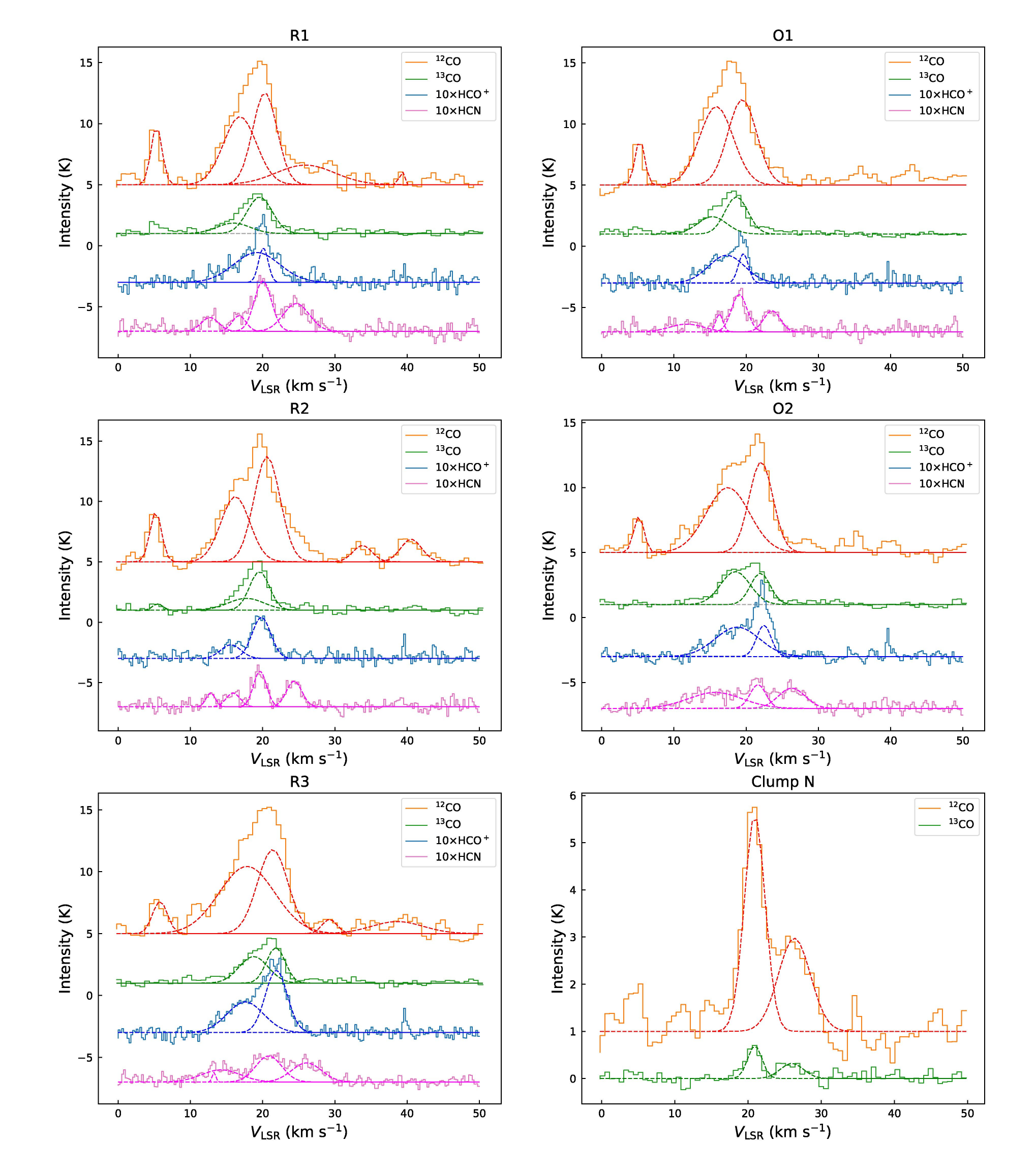}
    \caption{Averaged spectra and multi-Gaussian fitting results in velocity range from 0\,km\,s$^{-1}$ to +50\,km\,s$^{-1}$ for different molecular species. The intensity of \HCOp\ and HCN are multiplied by ten for comparison. 
    }
    \label{fig:all lines}
\end{figure*}

\begin{table*}
    \centering
    \caption{Multi-Gaussian fitting results and column densities for the secondary components of molecular lines as shown in Figure~\ref{fig:all lines}.}
    \label{tab:fit and column density}
    \begin{tabular}{ccccccc}
         \hline
         \hline
         Region      &  Molecule & $v_0$\,/\,km\,s$^{-1}$ & $T_{\rm peak}$ / K & FWHM / km\,s$^{-1}$ & Integration\,/ K \,km\,s$^{-1}$ & $N$(species) / cm$^{-2}$ \\
         \hline
                   &  \twCO\   & 16.86$\pm$1.28  & 5.07$\pm$ 2.18 &  
                    3.35$\pm$0.50 
                    & 24.30 $\pm$ 13.97  
                    & 2.20 $\pm$ 0.80 $\times10^{16}$\\
         R1        &  \thCO\   & 15.97$\pm$1.84  & 0.85$\pm$0.15 &  
                    3.57$\pm$1.3 
                    & 4.91 $\pm$ 1.14 
                    & 4.69 $\pm 1.09 \times10^{15}$\\
                   &  \HCOp\     & 19.18$\pm$0.20  & 0.24$\pm$0.03 &
                    4.19$\pm$1.08 
                    & 1.94 $\pm$ 0.43 
                    & 7.85 $\pm 1.50 \times10^{12}$\\
                   &  HCN         & 16.73$\pm$0.27  & 0.13$\pm$0.02 &
                    2.35$\pm$0.35 
                    & 0.75 $\pm$ 0.13
                    & 1.73 $\pm 0.73\times10^{12}$\\
        \hline
                   &  \twCO\   & 16.46$\pm$0.69  & 5.98$\pm$1.15 &
                    3.33$\pm$0.71 
                    & 21.07 $\pm$ 7.22  
                    & 5.06 $\pm 2.04 \times10^{17}$\\
         R2        &  \thCO\   & 17.72$\pm$1.92  & 1.00$\pm$0.43 &  
                    3.72$\pm$1.82 
                    & 6.15 $\pm$ 2.31 
                    & 5.92 $\pm 2.23 \times10^{15}$\\
                   &  \HCOp\     & 15.91$\pm$0.2  & 0.11$\pm$0.02 &
                    2.78$\pm$0.35 
                    & 0.40 $\pm$ 0.17 
                    & 1.31 $\pm 0.55 \times10^{12}$\\
                   &  HCN         & 15.90$\pm$0.19 & 0.11$\pm$0.02 &
                    2.35$\pm$0.28 
                    & 0.27 $\pm$ 0.09 
                    & 1.70 $\pm 0.55 \times10^{12}$\\
        \hline
                   &  \twCO\   & 17.88$\pm$1.63  & 5.26$\pm$1.56 &
                    4.63$\pm$1.82 
                    & 52.61 $\pm$ 19.69 
                    & 8.56 $\pm 4.58 \times10^{16}$\\
         R3        &  \thCO\   & 18.76$\pm$0.88  & 2.16$\pm$0.35 &  
                    3.33$\pm$1.33 
                    & 10.83 $\pm$ 3.25 
                    & 1.03 $\pm 0.30 \times10^{16}$ \\
                   &  \HCOp\      & 17.60$\pm$0.54  & 0.25$\pm$0.02 &
                    3.73$\pm$0.28 
                    & 1.57 $\pm$ 0.32 
                    & 5.12 $\pm 1.06 \times10^{12}$\\
                   &  HCN         &                 & 
                   ({\it undistinguishable}) &  \\
        \hline
                   &  \twCO\   & 15.86$\pm$2.75  & 6.42$\pm$1.04 &
                    3.65$\pm$0.82 
                    & 31.82 $\pm$ 22.19 
                    & 6.43 $\pm 4.2 \times10^{16}$\\
         O1        &  \thCO\   & 15.30$\pm$1.58  & 1.43$\pm$0.55 &  
                    3.33$\pm$1.05 
                    & 7.17 $\pm$ 3.21 
                    & 6.96 $\pm 2.48 \times10^{15}$\\
                   &  \HCOp\     & 17.29$\pm$0.28  & 0.024$\pm$0.03 &
                    3.71$\pm$0.14 
                    & 1.41 $\pm$ 0.20 
                    & 4.59 $\pm 0.64 \times10^{12}$\\
                   &  HCN         &                 & 
                   ({\it undistinguishable})  &  \\
        \hline
                   &  \twCO\   & 17.57 $\pm$0.97  & 5.67$\pm$0.57 &
                    4.08$\pm$1.45 
                    & 37.60 $\pm$ 12.53 
                    & 7.04 $\pm 2.72 \times10^{16}$\\
         O2        &  \thCO\   & 18.50$\pm$ 0.58 & 2.53$\pm$0.23 &  
                   3.63$\pm$1.29  
                   & 12.94 $\pm$ 3.2  
                   & 1.23 $\pm 0.3 \times10^{16}$\\
                   &  \HCOp\     & 18.78$\pm$0.39  & 0.22$\pm$0.01 &
                    4.10$\pm$0.19 
                    & 1.65 $\pm$ 0.21  
                    & 5.37 $\pm 0.77 \times10^{12}$\\
                   &  HCN         &                 & 
                   ({\it undistinguishable}) & \\
         \hline
                  &  \twCO\   & 25.99$\pm$0.35  & 1.98$\pm$0.16 &
                    3.35$\pm$1.05 
                    & 10.90 $\pm$ 2.34 
                    & 1.14 $\pm 0.26 \times10^{16}$ \\
         Clump N  &  \thCO\   & 26.04$\pm$ 0.34 & 0.32$\pm$0.06 &  
                   2.98$\pm$1.02  
                   & 1.30 $\pm$ 0.48  
                   & 1.46 $\pm 0.56 \times10^{15}$\\
        \hline
    \end{tabular}
    \begin{tablenotes}
        \footnotesize
        \item Note------$v_0$ denotes the center velocity of the secondary components. $N$ in the 7th columns denotes the column density of the specific molecular species.
    \end{tablenotes}
\end{table*}

\begin{table*}
    \centering
    \caption{Calculated parameters for the secondary components of molecular lines as shown in Figure~\ref{fig:all lines}.}
    \label{tab:column density and mass}
    \begin{tabular}{cccc}
         \hline
         \hline
         Region     &  $N$(\HCOp)/$N$(CO) 
         & $M$ /$\Msun$  & $\nHH$ / cm$^{-3}$\\
        \hline
         R1        &  3.6 $\pm 1.6\times10^{-4}$  & 
                  1800 & 300 \\
        \hline
         R2        &  2.6 $\pm 1.5\times10^{-5}$ & 
                    2200 & 320 \\
        \hline
         R3        &  6.0 $\pm 3.4\times10^{-5}$ & 
                    3900 & 600 \\
        \hline
         O1         & 7.0 $\pm 4.6\times10^{-5}$ & 
                    1700 & 400 \\
        \hline
         O2        &  8.0 $\pm 3.1\times10^{-5}$ & 
                   7000 & 500 \\
         \hline
         Clump N   & - - - & 1500 & 40 \\
        \hline
    \end{tabular}
    
\end{table*}

Figure~\ref{fig:gridmap} is a grid map of \HCOp\ line spectra at $\VLSR$ between $0\km\ps$ and $+50\km\ps$. 
Inside the radio boundary of the SNR, there is little line features; 
by contrast, evident line features are present outside the boundary, also illustrating that there is a sharp interface between the SNR and the outer molecular gas. 
Notably, in the outer region, the line features around $\sim +20\km\ps$ are asymmetric, extending from $\sim+10\km\ps$
to $\sim+25\km\ps$.

Specifically, as shown in Figure~\ref{fig:gridmap}, we select three on-boundary regions, R1, R2, and R3, along the southeastern border of the remnant and two outer regions, O1 and O2, for the use of extraction of molecular line spectra around the SNR. The five selected regions do not overlap with the HII regions discussed in \citet{Paron_kes67_clump, paron_south_clump}. 
The data of the CO lines and the data of the \HCOp\ and HCN lines come from different telescopes; for uniform format, the CO data have been regridded to the coordinate and spatial resolution of the \HCOp\ and HCN data when extracting the molecular spectra.

Figure~\ref{fig:all lines} shows the averaged emission line spectra of the four molecular species HCO$^{+}$, HCN, \twCO, and \thCO\ extracted from the five selected regions.
The line spectra of Clump N are extracted from the region marked in Figure~\ref{fig: fig1_cont}.
In the spectra of the three regions along the SNR boundary\,(R1, R2, and R3), we see small component of \twCO\ at $\sim$+5\,km\,s$^{-1}$ that corresponds to the local gas emission \citep{Paron_kes67_clump}. 
For all of the four molecular species in regions R1, R2, and R3, main components appear around $+20$\,km\,s$^{-1}$ with a secondary component from $\sim$+10\,km\,s$^{-1}$ to $\sim$+18\,km\,s$^{-1}$.
The Gaussian fitting results of the secondary components are given in Table~\ref{tab:fit and column density}. 
The secondary components could represent the blue-shifted broadened line wings as a signature of disturbance of the molecular gas by passing shock waves \citep[e.g.,][]{Jiang}. 
However, in Figure~\ref{fig:gridmap}, the grid map of \HCOp\ shows that similar line profiles are also seen in the outer regions O1 and O2, which indicates that the secondary components in the line profiles do not appear only along the southeastern boundary of the SNR.

\begin{figure}
	\includegraphics[width=0.5\textwidth]{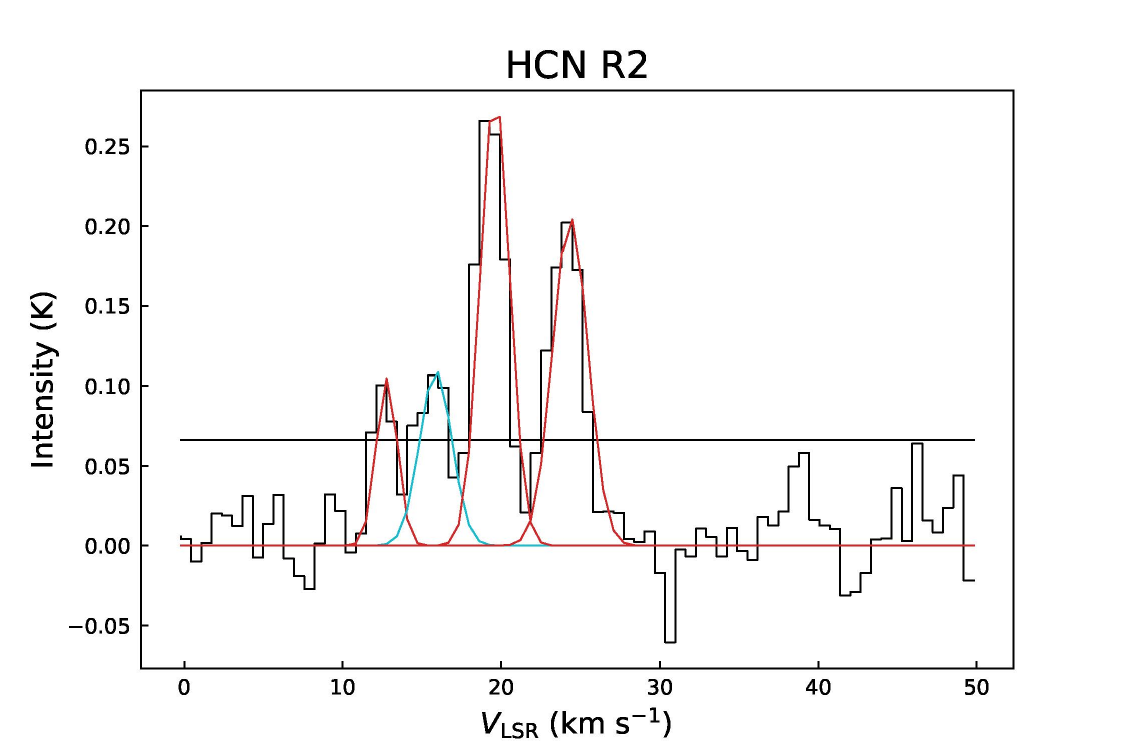}
    \caption{The line profiles of HCN in region R2, with 3\,$\sigma$ line in black. The red components show the fitting results of HFS with the $F$=1-0 line, which centered at $\sim$19.63\,km\,s$^{-1}$. The component of HCN fitted in cyan coincides with the secondary component of HCO$^{+}$ in region R2 and is not related to the HFS lines. The velocity resolution has been rebinned to 0.65$\km\ps$ for better displaying effect.}
    \label{fig:HCN_R2}
\end{figure}

Furthermore, the line profiles of HCN comprise at least three components in the range of  $\sim$
+10\,km\,s$^{-1}$ to $\sim$+30\,km\,s$^{-1}$ (see Figure~\ref{fig:all lines}). 
The HCN rotational levels include hyperfine structure (HFS) and have three components. Each hyperfine level is designated by a quantum number $F$ (= $I$ + $J$) and the strengths of the HCN (\Jotz) HFS lines is 1:5:3 in the optical thin limit. 
Theoretically, the LSR velocity difference between the $F =\,0$\,--\,1 line and $ F =\,2$\,--\,1 line is $-$7.1\,km\,s$^{-1}$, and that between the $F =\,1$\,--\,1 line and $F =\,2$\,--\,1 line is 4.9\,km\,s$^{-1}$ \citep{HCN_HFS}. 
In Figure~\ref{fig:HCN_R2}, we show the fitting result taking the HCN spectrum of region R2 as an example, which includes four components: those fitted in red are related to HFS and that fitted in cyan is related to a secondary or broadened component. 
The $F =\,2$\,--\,1,  $F =\,0$\,--\,1, and $F =\,1$\,--\,1 lines are centered at +19.63\,km\,s$^{-1}$, +12.79\,km\,s$^{-1}$, and +24.35\,km\,s$^{-1}$, respectively. The separation  between the $F =\,0$\,--\,1 and $F =\,2$\,--\,1 lines is $-$6.84\,km\,s$^{-1}$, and that between the $F =\,1$\,--\,1 and $F =\,2$\,--\,1 is 4.72\,km\,s$^{-1}$. 
Although the measured separations are slightly different from the theoretical values, the HFS of the HCN molecules can be identified because the differences are similar to or less than the velocity resolution (0.25$\km\ps$). 
The secondary or broadened component of HCN is centered at 15.90\,km\,s$^{-1}$, which could be of the same origin as the secondary or broadened components of \twCO\ and \HCOp\ in region R2.

For Clump N, we extract the averaged spectrum of two components of the line profiles, also shown in Figure~\ref{fig:all lines}. Besides the main component of \twCO\ in $\VLSR$ range $\sim+18 \km\ps$ -- $\sim +24\km\ps$, there is also a secondary component in $\VLSR$ range  $\sim+24 \km\ps$ -- $\sim +30\km\ps$. The Gaussian fitting results for Clump N are also listed in Table~\ref{tab:fit and column density}. 

\begin{figure}
	\includegraphics[width=0.5\textwidth]{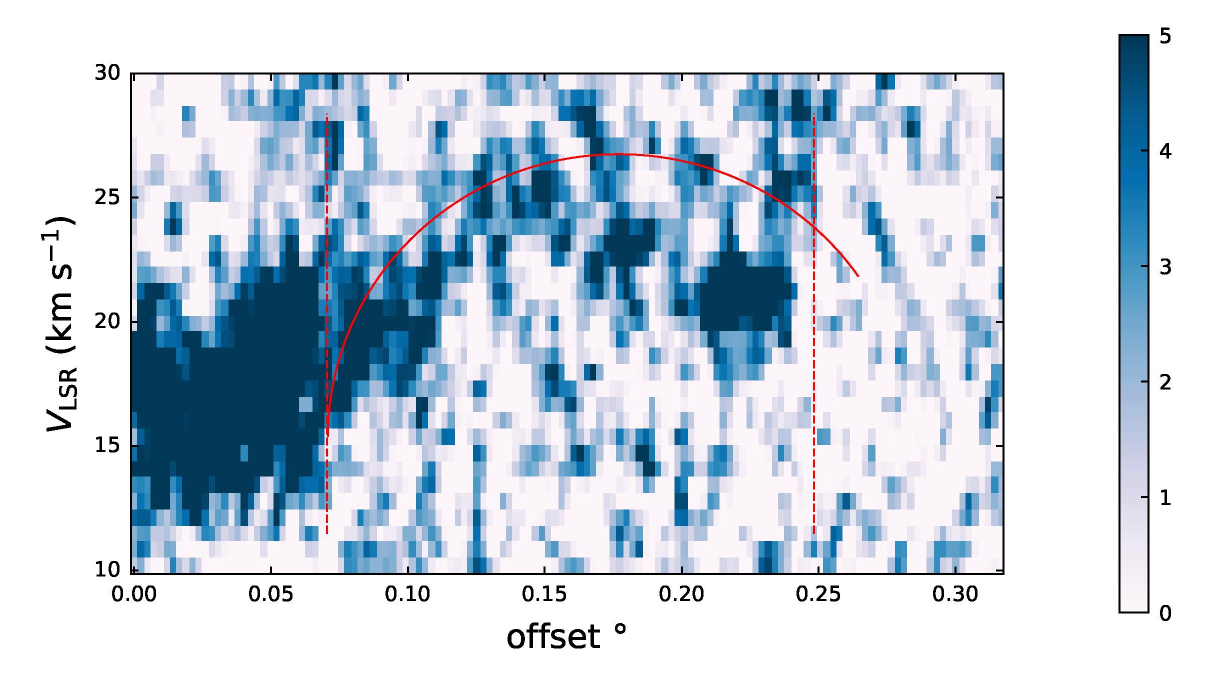}
    \caption{The position-velocity diagram extracted along the path (marked in red in Figure~\ref{fig:velocity_channel_12CO}) from region R2 to Clump N. The red vertical dashed lines delineate the position of region R2 and Clump N (marked by the orange diamonds in Figure~\ref{fig:velocity_channel_12CO}). The red arc indicates a general P-V pattern.}
    \label{fig:pv_diagram}
\end{figure}

We produce a position-velocity (P-V) diagram Figure~\ref{fig:pv_diagram} along the red line from region R2 to Clump N marked in Figure~\ref{fig:velocity_channel_12CO}. 
Figure~\ref{fig:pv_diagram} shows an arc-like distribution pattern, which indicates a $\VLSR$ span from $\sim +14\km\ps$ to $\sim+26\km\ps$. Such a pattern suggests an expanding gas structure centered at $\VLSR\sim+20\km\ps$.

\subsubsection{Parameters of Molecular Gas}
Using the local thermodynamic equilibrium (LTE) hypothesis, we calculate the column densities of \twCO, \thCO, \HCOp, and HCN molecules in the secondary or broadened components for the molecualar gas in the five selected regions and Clump N and list the results in Table~\ref{tab:fit and column density}. Here we have assumed the \twCO\ line is optically thick while the other lines of \thCO, \HCOp, and HCN are optically thin, and thus the column densities for the optically thick and thin cases are given by \citep{Calculate_column_dencity}: 

\begin{equation}
\begin{aligned}
    N_{\rm thick} & =  \frac{3h}{8 \pi^3 \mu^2 J_\mu R_{\rm i}} 
    \left(\frac{k T_{\rm ex}}{hB_{\rm 0}}+\frac{1}{3}\right)
    \frac{\exp{(E_{\rm u}/kT_{\rm ex}})}{\exp{(E_{\rm u}/kT_{\rm ex})} - 1 }\\ & \times \int -\ln{\left[1-\frac{T_{\rm R}}{J_{\nu}\left(T_{\rm ex}\right)-J_{\nu}(T_{\rm bg})}\right]}dv 
\end{aligned}
\end{equation}
and
\begin{equation}
\begin{aligned}
    N_{\rm thin} & = \frac{3h}{8 \pi^3 \mu^2 J_\mu R_{\rm i}}\left(\frac{k T_{\rm ex}}{hB_{\rm 0}}+\frac{1}{3}\right) \exp{\frac{E_{\rm u}}{kT_{\rm ex}}} \\ & \times\left(\exp{\frac{h\nu}{kT_{\rm ex}}}-1\right)^{-1}\int\tau_{\rm \nu}dv ,
	\label{equ:HCO+}
\end{aligned}
\end{equation}
respectively, where $J\left(T\right)=h\nu/[\exp{(h\nu/kT)}-1]$, and $T_{\rm bg}$ is the cosmic microwave background temperature (also see \citet{Calculate_column_dencity} for definition of other symbols in the above equations).
The optical thin assumption for \HCOp\ may underestimate the column density.

The excitation temperature of \twCO\ is estimated using \citep{Calculate_column_dencity}
\begin{equation}
\begin{aligned}
    T_{\rm ex} & = \frac{h\nu/k_{\rm B}}{\ln{\left[1+\frac{h \nu/k_{\rm B}}{T_{\rm max}+ J_{\rm \nu}(T_{\rm cmb)}}\right]}} \sim 14\K,
\end{aligned}
\end{equation}
where $k_{\rm B}$ is the Boltzmann constant, and $T_{\rm max}$ is the maximum brightness temperature taken from the molecular lines. 
Then, we calculate the column density ratios $N$({\HCOp})/$N$(\twCO), which are listed in Table~\ref{tab:column density and mass}.
We obtain the column densities of hydrogen molecules from $N({\rm ^{13}CO})$ (as listed in Table~\ref{tab:fit and column density}) using abundance ratio $N({\rm H_2 })/N({\rm ^{13}CO}) \approx 7 \times 10^5$ \citep{13CO-to-H2=7e5}. 
On the assumption that the line-of-sight depth in each region is similar to the transverse size,
we estimate the mass of the secondary component by $M = 2.8m_{\rm H}N({\rm H}_{\rm 2}){\cal A}$ (where ${\cal A}$ is the projected area for each region from which the molecular line spectra are extracted (see Figure~\ref{fig: fig1_cont} and Figure~\ref{fig:gridmap})) and number density $\nHH$.
The mass and number density values are listed in Table~\ref{tab:column density and mass}. 
The lower limit of the mass of the secondary component in the molecular belt is estimated by summing the masses of the secondary components in the five selected regions. The upper limit is estimated by multiplying the average column density of the five regions by the projected size of the belt.
The estimated mass ranges from $\sim 1.6\E{4}$ to $\sim4.7\E{4} \Msun$, and we will use the average $\sim 3.1\E{4} \Msun$ in later calculations.

\subsection{{\sl Fermi}-LAT Data Analysis}\label{sec:fermi}
\subsubsection{Spatial Analysis}

\begin{figure*}
	\includegraphics[width=1.\textwidth]{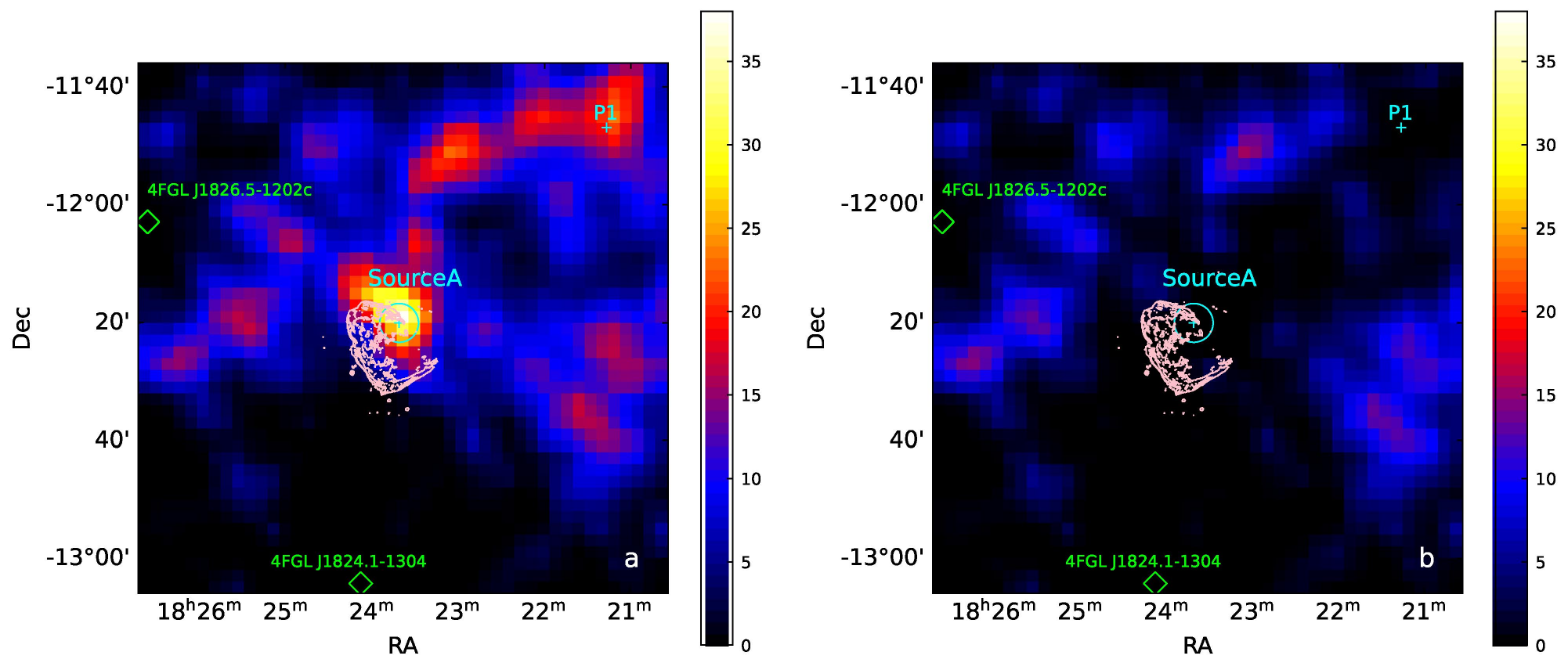}
    \caption{TS maps of 1.5$^{\circ}$$\times$1.5$^{\circ}$ field covering SNR Kes\,67 in the energy range of 0.2--500\,GeV. The image scale of the TS maps is 0.05$^{\circ}$ per pixel. The radio contours 
    are the same as those in Figure~\ref{fig: Integerated maps}}. 
    The cyan circle marks the best fitted location of `Source A' and the range of 95\% positional uncertainty.
    Left: TS map without source points designated. Right: residual TS map after adding the point source `Source A' and source `P1'.
    \label{fig:tsmap}
\end{figure*}

We use the 4FGL-DR4 catalogue in our \gray\ data analysis and there are no catalogue sources in the field of view around SNR Kes\,67. 
When fitting the models, we free the spectral parameters of sources within 5$^\circ$ ROI above 5$\sigma$ and the normalization of the Galactic and isotropic diffuse background. 
Then we fix all parameters except the normalization parameters of the Galactic and isotropic diffusion background component and generate the residual test-statistic\ (TS) map of the 1.5$^{\circ}$$\times$1.5$^{\circ}$ region centered at the SNR. 
The test statistic is defined as TS$ = 2\log{\left({\cal L}_1/
{\cal L}_0\right)}$, in which ${\cal L}_0$ is the maximum likelihood of the null hypothesis and ${\cal L}_1$ is the maximum likelihood with a putative source located in the tested pixel. 
The TS map is shown in Figure~\ref{fig:tsmap}. 
As seen in Figure~\ref{fig:tsmap}a, there is residual $\gamma$-ray emission in the north of the SNR. 

We select photon events above 1 GeV for spatial analysis to reduce uncertainties. We add a source with a single power-law (PL) spectrum into the background model at the pixel of the maximum TS value, which is denoted as `Source A'.
With the \texttt{extension} method in {\small Fermipy}, we do not detect any extension for `Source A', 
so we still use the point source (PS) model for further analysis. 
Using the \texttt{source\,\,localization} method in {\small Fermipy}, we obtain the best fitted position of `Source\,A' , 
which is (R.A.=$275.9203^{\circ}$, Dec.=$-12.3349^{\circ}$, J2000), with the 95\% positional uncertainty is 0.059$^{\circ}$, as illustrated in Figure~\ref{fig:tsmap} (and also marked in Figure~\ref{fig: Integerated maps}c).
This point source could explain the emission near the SNR and there is still residual emission in the region. 
We add another point sources (`P1'), as listed in Table~\ref{tab:4sources} with a single PL spectrum to the background template. 
Then we test the two point sources template in the energy range of 0.2--500 GeV. 
The final residual map is shown in the right panel of Figure~\ref{fig:tsmap} in  0.2--500 GeV, there is almost no residual \gray\ emission with the two additive sources. 

\begin{table}
    \centering
    \caption{Coordinates for two sources in energy range of 0.2--500 GeV}
    \label{tab:4sources}
    \begin{tabular}{lccr}
         \hline
         \hline
         Name      &  R.A.(J2000) & Dec.(J2000)  & TS\\
         \hline
        SourceA   &  275.9203   & $-$12.3349 & 52.83 \\
        P1   &  275.3207     & $-$11.7820   &  26.64\\
         \hline
    \end{tabular}
\end{table}

\subsubsection{Spectral Analysis}
With the PS spatial model for `Source A', we select photon events in the energy range of 0.2--500\,GeV for $\gamma$-ray spectral analysis. 
In the spectral analysis, we also take into account three other spectral models: log-parabola (LP), exponentially cutoff power-law (ECPL), and broken power-law (BPL) in addition to the PL model. 
The formulae of these spectral models are listed in Table~\ref{tab:formulae}. 
To determine the best fitted model, we define the TS$_{\rm spec}$ = 2$\log{\left(\mathcal{L}_{\rm spec}/\mathcal{L}_{\rm PL}\right)}$and choose the model with the largest TS$_{\rm spec}$ value. 
The test results are listed in Table~\ref{tab:spectral select}, and thus we choose the PL model. 

\begin{table}
    \centering
    \caption{Formulae for $\gamma$-ray spectra}
    \label{tab:formulae}
    \begin{tabular}{lccr}
    \hline
    \hline
    Name &  Formula & Free Parameters\\
    \hline
     PL  & $dN/dE = N_0\left(E/E_0\right)^{-\Gamma}$ 
         & $N_0, \Gamma$ \\
     ECPL& $dN/dE = N_0\left(E/E_0\right)^{-\Gamma}\exp{\left(-      E/E_{\rm cut}\right)}$ 
         & $N_0, \Gamma, E_{\rm cut}$ \\
     LogP& $dN/dE = N_0\left(E/E_0\right)^{-\Gamma-\beta\log{\left(E/E_0\right)}}$ 
         & $N_0, \Gamma, \beta$ \\
     BPL & $dN/dE = N_0\begin{cases}
                    \left(E/E_{\rm b}\right)^{-\Gamma_1}& \text{$E\leq E_{\rm b}$}\\
                    \left(E/E_{\rm b}\right)^{-\Gamma_2}& \text{$E\geq E_{\rm b}$}
                    \end{cases}$
         & $N_0, E_{\rm b}, \Gamma_1, \Gamma_2$ \\
     \hline
    \end{tabular}
\end{table}

\begin{table}
    \centering
    \caption{Test results of spectral models for `Source A'}
    \label{tab:spectral select}
    \begin{tabular}{lr}
    \hline
    \hline
         Spectral Model & TS$_{\rm spec}$\\
    \hline
         PL             &  0  \\
         ECPL           &  $-$1.03  \\
         LogP           &  $-$1.03  \\
         BPL            &  $\sim\num{-1.5e6}$  \\
    \hline
    \end{tabular}
\end{table}

{\small Fermipy} gives the TS value of `Source A' is 52.83 with the PS spatial model and the PL spectral model, which is 6.5$\sigma$. We obtain the 0.2--500\,GeV \gray\ flux $\sim\num{9.6e-12}\erg\cm^{-2}\ps$ and the index $\Gamma=2.35 \pm 0.11$, which leads to the luminosity $\sim 1.78\E{34} {\rm erg}\ps$ at a distance of $d\sim$14\kpc.

\begin{figure}
    \includegraphics[width=0.5\textwidth]{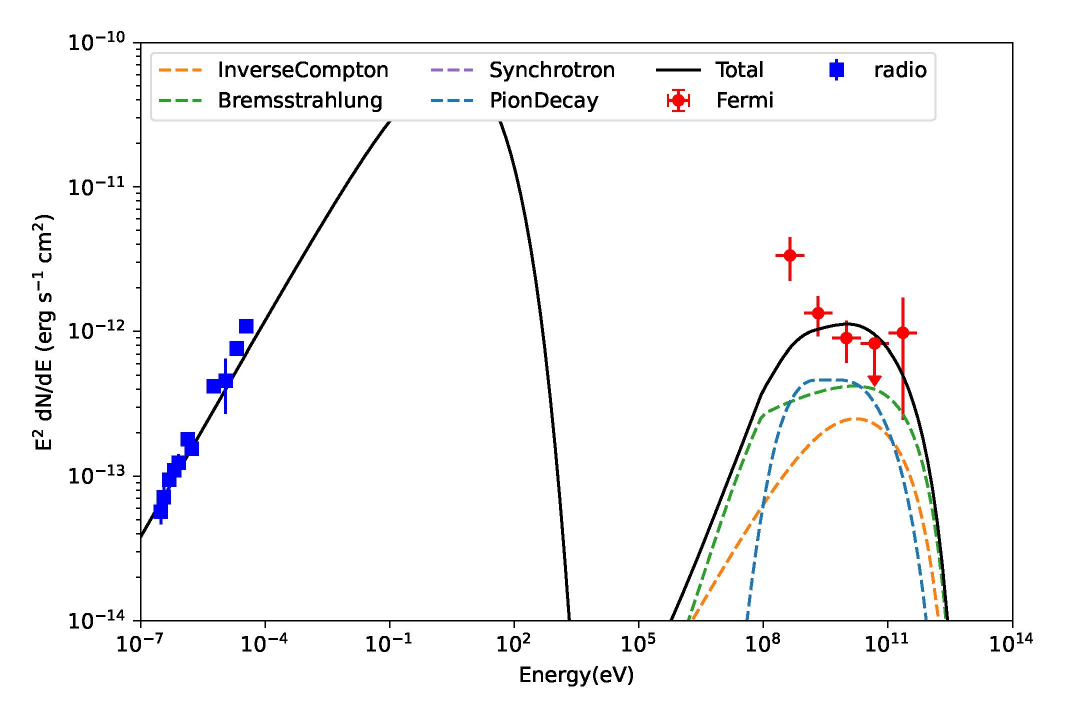}
    \caption{The SED of `Source A'. The radio data are listed in Table~\ref{tab:radio data}.}
    \label{fig:sed}
\end{figure}

Using the \texttt{SED} method of {\small Fermipy}, the spectral energy distribution\,(SED) of `Source A' in the  energy range of 0.2--500\,GeV is generated by using the maximum likelihood analysis in five logarithmically spaced energy bins. 
In the fitting process, we free the normalization parameters of the sources within 3$^{\circ}$ from the ROI center and the Galactic and isotropic diffuse background parameters. In the energy bins, when the TS value of `Source A' is less than 4, we calculate the 95\% confidence level upper limit of flux. The SED results are shown in Figure~\ref{fig:sed}. For further restriction, we use the radio flux data presented in the previous literature (listed in Table~\ref{tab:radio data}) of the entire SNR.

\section{Discussion}\label{sec:discussion}

\subsection{The SNR-MC Association}

As revealed in our spatial analysis of the molecular environment of SNR \snr, the molecular belt that closely surrounds the southeastern boundary of the SNR is recessed in the band-like molecular gas structure (\S\ref{sec:molecular distribution}). 
The molecular belt looks like a cushion between the SNR and the molecular band. 
This is a perfect morphological evidence for the physical contact among the SNR, the belt, and the band. 
Actually, the elongated structure (named `feature A') discerned  by \citet{Dubner1999} in a much lower spatial resolution seems similar to the belt our study shows and was suggested as the indication of SNR-MC physical association. 
Figure 1 in \citet{Paron_kes67_clump} for \thCO\ (\Jotz) emission with similar angular resolution also presents a comparable belt-like morphology.

\subsubsection{The asymmetric broad line profiles}

The SNR-MC association is supported by the asymmetric molecular line profiles that we obtain (\S\ref{sec:lines}), which could be caused by the perturbation from SNR shock or progenitor's wind.
The secondary components at $\sim+16\km\ps$ of the four molecular species in the five regions in the belt (see Table~\ref{tab:fit and column density}) could result from the perturbation that the molecular gas suffers.
If the perturbation come from the SNR shock, the asymmetric broad line profiles in the positions on the SNR shock front (such as regions R1, R2, and R3) could be explained.
However, the line profiles in the positions away from the SNR shock front (such as regions O1 and O2) are very similar to those in regions R1, R2, and R3, and cannot be ascribed to SNR shock impact.
Notwithstanding, the asymmetric broad line profiles in the molecular belt can be explained if the source of perturbation is the wind of the SNR's progenitor star,
which could drive a wind shock into the band-like structure.
In the latter case, the belt may be a segment of shell swept up by the wind.
This segment, with a mass $\sim3.1\E{4}\Msun$ 
was deposited a kinetic energy $\sim1\E{49}\,(v_{\rm b}/6\km\ps)^2\erg$ (with $v_{\rm b}$ the velocity of the shell motion) from the wind before the impact by the supernova blast wave.

Also, from the view angle of shock chemistry, the HCN-to-\HCOp\ and \HCOp-to-\twCO\ abundance ratios seem more consistent with the scenario of the molecular belt's perturbation by the progenitor's wind than that by the SNR shock, as discussed below.

Moreover, for Clump N in the north, the secondary component around $\sim+26\km\ps$ can be ascribed to a red-shifted broadening from the main component around $\sim+20\km\ps$.
It is consistent with a patch of backward moving molecular gas also driven by the progenitor's wind.

Furthermore, the blue-shift in the molecular line wings in the belt and the red-shift in Clump N are consistent with the expanding motion at velocity $\sim6\km\ps$ that is uncovered by the P-V diagram (Figure~\ref{fig:pv_diagram}), which can be compatible with the case of wind-driven shell.

\subsubsection{The molecular abundance ratios}
\label{sec:shock code}
The abundance ratio $N$({\HCOp})/$N$(\twCO) in the five regions is in a range from $7.2\times10^{-5}$ to $4.0\times10^{-4}$. 
Similarly, we use the secondary or broadened components in region R2 (shown in Figure~\ref{fig:all lines} and Figure~\ref{fig:HCN_R2} ) to estimate the abundance ratio $N({\rm HCN})/N({\rm HCO}^+)$ in the region and obtain a value $\sim1.2$. 
This ratio value is very similar to the $N({\rm HCN})/N({\rm HCO}^+)$ ratio, $1.9\pm0.9$, which is obtained in the diffuse interstellar medium \citep{HCO+/HCN}. 
However, the ratio is not much different from those obtained in SNR IC\,433.
The ratios obtained in IC\,433 are model dependent, ranging from 2.7 to 30 \citep{IC443_shocked_gas}.
The HCN-to-\HCOp\ ratios can also be much lower in SNRs; 
for example, the ratio is $\sim0.4$ for the MC interacting with SNR G349.7+0.2 \citep{HCN_G349d7}.

For the possibility of shock-induced asymmetry in molecular line profiles, the column density ratio $N({\rm HCO}^+ )/N({\rm ^{12}CO})$ may reflect the chemical effect of the shock in two cases.
The first case is that a preshock gas like that in the belt with a density $\sim10^3\cm^{-3}$ is traversed by the shock driven by the SNR at a velocity 10\,--\,$50\km\ps$ for a timescale $\sim20$\,kyr. 
This case is assumed for the emission from regions R1, R2, and R3. 
Actually, the beam size of the PMOD observation of \HCOp\ and HCN is $\sim 1'$, the emission from R1, R2, and R3 may contain the contribution from the molecular gas disturbed by the SNR shock (if any).
For LECRs , the cross section of ionization can be estimated by 
$4\pi a_0^2 [0.71 \ln{(1 + m_{\rm e} E_{\rm p}/m_{\rm p} I({\rm H}))}+ 1.63] / (m_{\rm e} E_{\rm p} / m_{\rm p} I({\rm H})) $, 
in which $I({\rm H}) = 13.6$ eV, $a_0 = 0.539 $ \AA \,
\citep{cross_section_2009, cross_section_1989}. The cross section values of 10 MeV and 100 MeV CR protons are $\sim 3.8\E{-18}$ and $\sim 6.8 \E{-19}$ cm$^{-2}$, respectively, corresponding to the mean free path values of protons $\sim 1.3\E{14} $ and $7.3\E{14}$ $(\nHH/2000\cm^{-3})^{-1}\cm$, much less than 1 pc.
Therefore, 
the accelerated CR particles cannot pass through the molecular belt to arrive at the outer regions O1 and O2.

The second case is that a preshock molecular gas with a density $\sim10^2\cm^{-3}$ is swept up by the wind-driven shock 
having evolved to the steady state. This case is assumed for all the five regions in the molecular belt that is considered as a segment of wind-swept shell.

\begin{figure*}
    \includegraphics[width=1.0\textwidth]{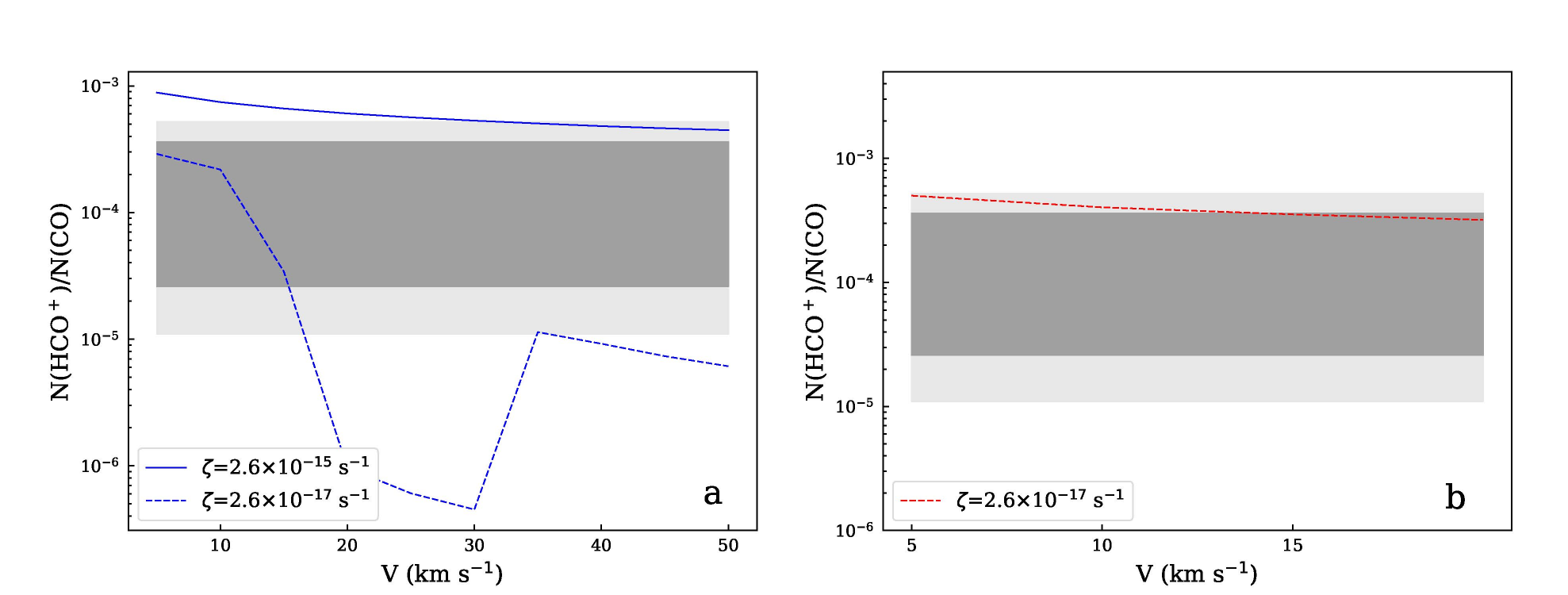}
    \caption{\HCOp-to-\twCO\ abundance ratio calculated using the Paris-Durham shock model, 
    in comparison with the abundance ratio ranges (in dark grey) obtained from observations with the best-fitted parameters listed in Table~\ref{tab:column density and mass} and the uncertainty ranges (in light grey).
    The used CR ionization rates ($\zeta$) are marked in the panels.
    {\sl a}) The ratio for a high-density ($10^3\cm^{-3}$) molecular gas.
    {\sl b})  Same as in a), but for lower-density ($10^2\cm^{-2}$) molecular gas.}

    \label{fig:PD shock code}
\end{figure*}

We apply the Paris-Durham shock model\footnote{\url{https://ism.obspm.fr/shock.html}} to examine the chemistry evolution on the path of shock propagation in these two cases \citep{shock_code_1,shock_code_2}.
The Paris-Durham chemistry network contains 140 molecular species and over 3000 chemical reactions, including \HCOp\ and CO.
Here, the \HCOp-to-\twCO\ abundance ratio is calculated for a range of velocity, $v_{\rm c}$, of shock propagation in the MC for each case.
In the model calculation, the magnetic field strength in the cloud is taken as $B\approx \sqrt{\nHH/\cm^{-3}}\,\mu\mbox{G}$. 
In addition, we take into account the molecular ionisation induced by low-energy ($<280$\,MeV) CRs.
Two CR ionization rates are adopted: $\zeta=\num{2.6e-17} \ps$ typical of Galactic value \citep{2.6e-17} and $\zeta=\num{2.6e-15}\ps$ for the gas ionized by SNR-accelerated CRs \citep[e.g.,][]{W51C_Ce, W28_Va, W49B, Tu_W28}.
The calculation results are shown in Figure~\ref{fig:PD shock code}.

For the first case assumed for SNR shock (Figure\,\ref{fig:PD shock code}a), the calculated abundance ratios for the five regions for $\zeta=2.6\E{-15}\ps$ is slightly above the range of observed values, but still within the uncertainty range.
Thus the chemical effects of the SNR shock together with CRs cannot be excluded. 
For $\zeta=2.6\E{-17}\ps$, 
which represents that the ionizing CRs from the SNR are not at work, 
the calculated ratios can match the observational values at $v_{\rm c}$ around $10\km\ps$ for a C-type shock \citep{C-type_shock}.
For $v_{\rm c}$ increases above $\sim 20\km\ps$, the shock wave switches into J-type and tends to dissociate the molecules.

\citet{Fe} found enhanced Fe\,I K$\alpha$ line emission from SNR Kes\,67,  
which arises from a ``C''-shaped shell-like structure.
They suggested that the Fe\,I K$\alpha$ line should be emitted from the dense clouds near the LECR protons acceleration sites.
As also seen from Figure~\ref{fig: Integerated maps}c, this Fe\,I\,K$\alpha$ emitting shell does not appear to cover the southeastern molecular belt, 
leaving a lack of evidence of LECR proton acceleration traced by Fe\,I\,K$\alpha$ emission along the southeastern boundary of the SNR. 
This is unfavorable to the first case discussed above.

For the second case assumed for the progenitor's wind driven molecular belt (Figure\,\ref{fig:PD shock code}b), the calculated ratios appear consistent with the observation.

\subsubsection{Impact of the Progenitor's Wind}

A scenario of the molecular environment of SNR\,\snr\ can commonly be pointed to by the above various analysis
(spatial distribution of the molecular belt, asymmetric broad lines, P-V diagram of \twCO, and  abundance of \HCOp):
there is an incomplete molecular shell of bubble
with an expansion velocity $v_{\rm b}\sim6\km\ps$,
which was blown by the progenitor's wind and is interacting with the SNR.
The southwestern molecular belt and the northern Clump~N are parts of the materials of the shell.
Actually, wind-blown, expanding molecular shells/bubbles have also been seen in some other SNRs, such as Tycho's SNR \citep{Tycho_shell}, VRO 42.05.01 \citep{VRO_42.05.01}, G352.7$-$0.1 \citep{G352.7-0.1}, and G9.7$-$0.0 \cite{Tu_13SNRs_2024}.
For Kes 67, applying the wind bubble model by \citet{Castor_interstellar_bubble}, the timescale of the bubble evolution can be estimated from radius $\sim 7.5'$ or $\sim$30\,\parsec\ (at $\sim 14$ kpc):
\begin{equation}
  t_{\rm b}=\frac{3r_{\rm b}}{5v_{\rm b}}\sim3\E{6} \left(\frac{r_{\rm b}}{30\parsec}\right)\left(\frac{v_{\rm}}{6\km\ps}\right)^{-1}\yr 
\end{equation}
The kinetic luminosity of stellar wind is:
\begin{equation}
\begin{aligned}
    L_{\rm w} \sim1.7\E{37}  \left[\frac{n_0({\rm H}_2)}{10^2\cm^{-3}}\right]
     \left(\frac{r_{\rm b}}{30\parsec}\right)^2
    \left(\frac{v_{\rm b}}{6\km\ps}\right)^3\erg\ps,
\end{aligned}
\end{equation}
where $n_0({\rm H}_2)$ is the molecular number density ahead of the bubble shock.
A wind with such a high kinematic luminosity of order $10^{37}\erg\ps$ could be launched by a O-type (no later than O4) star with a wind velocity of $\gsim3\E{3}\km\ps$ in a duration of $t_{\rm b}\sim 3$\,Myr \citep{wind_bubble_2013}.
This timescale includes the Wolf-Rayet (WR) stage which could last $\sim5\E{5}$\,yr after the main-sequence (MS) age of the progenitor star \citep{WR_lifetime}.
For a WR star, the kinetic luminosity of the stellar wind can typically achieve 5$\E{37}\erg\ps$ with the mass loss rate $\sim 4\E{-5} \Msun$\,yr$^{-1}$ and stellar wind velocity $\sim 2\E{3}\km\ps$ \citep[e.g.,][]{WR_parameters}.

This dynamical evolution of the wind bubble is yet somewhat crude in ignoring the inhomogeneous distribution of the environmental gas, e.g., the density gradient from west to east, which may be responsible for the blowout morphology of the remnant as in the case of SNR Kes\,27 \citep{Kes27}.
 
The remarkable rectangular shape of SNR\,\snr\ in the southern edge is also seen in some other SNRs, such as Puppis\,A and 3C397.
\citet{rectangular_Puppis_A} reproduced such a shape by simulating the interaction between supernova blastwave and the bubble wall molded by the progenitor's magnetized bipolar winds.
The morphology of 3C397 was also suggested to possibly be related to a pre-supervova bipolar circumstellar environs  \citep{3c397}.
Therefore, it is possible that the wind of the \snr\ progenitor was bipolar and magnetized, inhibited by molecular gas of the belt in the south.

In \citet{Paron_kes67_clump}, the eastern clump, which is separated from the SNR radio shell boundary, 
contains a few infrared sources that are identified as protostars. 
The age of these young stellar objects ranges from 0.5 -- $10\E{5}\yr$.  
For the southern HII region studied in \citet{paron_south_clump}, the estimated age for the young massive star is $\sim1\E{5}$ yr.
The authors have discarded the possibility that the star formation was triggered by SNR \snr\ in comparison with the SNR age.
The smaller age estimate ($\sim 20$ kyr) in this work supports their judgement. 
However, in the scenario of progenitor's wind bubble that we inferred, the timescale of the wind-driven shell (or the molecular belt) $\sim 3$\,Myr is much larger than the timescales of the star formation given in \citet{Paron_kes67_clump, paron_south_clump}.  
Therefore, it is possible that the star formation was triggered by the expanding motion of the wind-driven shell. 

\citet{paron_south_clump} found that, for the southern clump, close to region R3 in this work, the virial mass is one order of magnitude larger than the mass calculated with LTE assumption, which indicates that the clump with this feature is not gravitationally bound. Thus, a source for external pressure $\sim 1.4\E{6} k_{\rm B} \cm^{-3}$K is needed to keep the clump bound. They suggested that the SNR shock front can play such a role.
However, as seen in Figure~3 in \citet{paron_south_clump}, the SNR shock front is not in contact or sufficient contact with this clump. 
Actually, the external pressure can be naturally provided in the scenario of stellar wind-driven bubble. 
The pressure in the bubble shell (or molecular belt) is similar to the ram pressure of the forward shock of the bubble
$\sim2.8n_0(\mbox{H}_2)\mH v_{\rm b}^2 
\sim 1.2\E{6} [n_0(\mbox{H}_2)/10^2\cm^{-3}] k_{\rm B} \cm^{-3}$K

\subsection{The Origin of Possible \gray\ Emission}
While there is hardly molecular ionization by SNR-accelerated LECR protons along the southeastern adjoining molecular belt (\S\ref{sec:shock code}) and there is signature of LECR protons that yield ``C''-shaped Fe\,I\,K$\alpha$ line emission in the interaction with dense gas \citep{Fe},
signs of high-energy protons that SNR \snr\ accelerates may have emerged in our studies. 
As shown in Figure~\ref{fig:tsmap}, the possible GeV \gray\ point source `Source A' is projectionally located in the north of the SNR, with the 68\% location uncertainty circle covering the Clump N at the same $\VLSR$ ($\sim+20\km\ps$) as that of the southeastern adjoining molecular belt (see Figure~\ref{fig: Integerated maps}c and Figure\,\ref{fig:velocity_channel_12CO}). 
The best-fitted position of `Source A' is coincident with the small region in yellow with radio index $\sim$ $-$0.3 (Figure 2b in \citet{low_freq_flux}). 
We have checked the SIMBAD Astronomical Database
 \citep{SIMBAD} within the radius of 2$\sigma$ location uncertainty of `Source A' and find 83 variables (or candidates), 8 stars, 1 AGB star, 1 red giant star, 1 emission object, 2 radio sources, 6 sub-millimeter sources, and 3 YSO candidates.
Although we cannot exclude the origin of $\gamma$-rays from YSOs or some kind of emission object, it is more natural to ascribe the \gray\ emission to the SNR-MC association system.

\begin{table}
    \centering
    \caption{Details of radio data.}
    \label{tab:radio data}
    \begin{tabular}{lcr}
    \hline
    \hline
        Frequency(GHz) & Flux(Jy) & Reference\\
    \hline
    0.074      & 76.2$\pm$13.8  & \citet{low_freq_flux} \\
    0.088      & 81$\pm$17      & \citet{low_freq_flux} \\
    0.118      & 80$\pm$14      & \citet{low_freq_flux} \\
    0.155      & 71$\pm$11      & \citet{low_freq_flux} \\
    0.200      & 62$\pm$9      & \citet{low_freq_flux} \\
        0.327          & 55       & \citet{radio_327MHz} \\
        0.408          & 38       & \citet{radio_408MHz} \\
        1.4            & 29.9$\pm$0.3  & \citet{Dubner1} \\
        2.7            & 17$\pm$7      & \citet{radio_2.7GHz} \\
        5              & 15.3$\pm$0.9  & \citet{radio_5GHz}  \\
        8.4            & 12.9$\pm$1.0  & \citet{radio_8.4GHz} \\
    \hline
    \end{tabular}
\end{table}

To analyze the possible origin of the $\gamma$-rays, we fit the broadband emission spectrum with both hadronic and leptonic processes considered.
We assume that the particles accelerated by the SNR shock have a PL form with a high-energy energy cutoff:
\begin{equation}
\begin{aligned}
    dN_i/dE =A_i (E_i/ 1\ {\rm GeV})^{-\alpha_i} \exp(-E_i/E_{{\rm cut},i})
\end{aligned}
\end{equation}
where $i={\rm e, p}$; $E_i$, $\alpha_i$, and $E_{{\rm cut},i}$ are the particle energy, the PL index, and the cutoff energy, respectively. The normalization $A_i$ is determined by the total energy above 1GeV that is converted from the explosion energy $E_{\rm SN}$ with a fraction of $\eta$.
In the calculation, we employ the parameter $K_{\rm ep} = A_{\rm e}/A_{\rm p}$ to control the ratio of electrons and protons.

To calculate the broadband SED, we consider four radiation mechanisms integrated in the PYTHON package Naima \citep{naima}: synchrotron \citep{syn}, inverse Compton \citep[IC,][]{ICS}, non-thermal bremsstrahlung \citep{bremsstrahlung}, and pion-decay \citep{pp_decay} processes.
For the IC process, the infra-red (IR) photons with a temperature of 35~K and an energy density of 0.6\,eV\,cm$^{-3}$ estimated from the interstellar radiation field \citep{seed_photons2} are also considered besides the cosmic microwave background (CMB). 
Here, MC Clump N is considered to be in contact with a fraction ($\sim1/6$) of the SNR shock surface to fit the entire SNR.
For the non-thermal bremsstrahlung and pion-decay processes, therefore, the average number density (averaged over the entire shock surface) of the target protons (with that the high energy particles interact)
is 
$\sim[2\nHH]/6\sim10\cm^{-3}$. 

In the SED fitting, we keep $\alpha_{\rm e}=\alpha_{\rm p}$ and adopt the typical explosion energy $E_{\rm SN}=10^{51}$~erg.
Due to the large number density, the contribution to the GeV $\gamma$-rays from the IC process can be ignored compared with the non-thermal bremsstrahlung.
Then, the GeV $\gamma$-ray spectrum mainly depends on $K_{\rm ep}$.
As shown in Figure~\ref{fig:sed}, the contribution by the non-thermal bremsstrahlung and the pion-decay processes are comparable for $K_{\rm ep}=0.1$.
To explain the data, $\alpha=2.0$, $\eta=0.04$, $B=80\ {\rm \mu G}$, and $E_{\rm cut, e,p}=1$~TeV are obtained.
For $K_{\rm ep}\gg0.1$ (resulting in $B\ll80\ {\rm \mu G}$), the leptonic $\gamma$-rays contributed mainly by the non-thermal bremsstrahlung dominate the GeV flux. 
With the magnetic field strength $B=80\ {\rm \mu G}$ and the SNR age of order $10^4$~yr, the synchrotron cooling break is about 0.5 TeV which is very close to the cutoff energy.
It means that the synchrotron cooling does not significantly shape the distribution of electrons for the lepton-dominated case.
On the other hand, for $K_{\rm ep} \ll 0.1$, the \gray\ flux are dominated by the pion-decay process. 
For Galactic CRs, the value of $K_{\rm ep} \sim 0.01$ is obtained by comparing the total electron and proton luminosities at Earth \citep[e.g.,][]{kep=0.01}. 
Considering the positional coincidence between the SNR and the molecular clump, `Source A' likely has a hadronic origin.
Therefore, in the SNR paradigm for the origin of the Galactic CRs, it is most likely that the \gray\ emission from `Source A' is powered by the SNR accelerated protons.

While the GeV $\gamma$-rays are detected in the location of `Source A' or Clump N, they are not detected in outer regions along the eastern and southern boundary, which are found rich in molecular gas.
This phenomenon, though, is not unusual in SNR-MC interaction systems. 
For example, SNR Kes\,69 is found to be associated with the 1720 MHz OH masers, including the compact masers found at the northeastern boundary \citep{1720MHz_masers} and extended masers along the southern boundary \citep{kes_69_extended_maser}. 
The 1720\,MHz OH masers are commonly regarded as robust signpost of the SNR-MC interaction.
CO and \HCOp\ line observations also demonstrate that the SNR is surrounded by dense molecular gas along both southern and northern boundaries \citep{Kes_69_zhou, Tu_13SNRs_2024}.
SNR HC\,40 are also revealed to be surrounded by molecular gas \citep[][and also see \citet{Jiang} for more references therein]{HC40_CO}.
Yet, there are no reports of GeV or TeV \gray\ emissions associated with these two SNRs.
In addition, SNRs Kes\,41 \citep{Kes41_LiuB}, 3C\,391 \citep{GeV_3c391} and G9.7-0.0 \citep[][Shen et al. in prep]{G9.7_GeV} are also embraced by molecular gas, but the \gray\ emissions associated with them are only detected towards certain corners.

There may be some potential physical reasons for 
\gray\ deficiency
in some portions of molecular gas surrounding the SNR boundary.
One possible reason is that the low energy particles are still trapped inside the SNR boundary and can not escape to illuminate the outside molecular gas in GeV band.
Based on the time-dependent escaping process \citep[the so-called $\delta$ escape,][]{gabici2009}, SNR Kes 67, at an estimated age of $10^4$~yr, can confine the particles with energies less than $\sim 300$~ GeV which is roughly consistent with the fitted cutoff energy.
Clump N seems to be embedded in a void of diffuse radio emission in the northern region by projection (see Figures \ref{fig:velocity_channel_12CO} and \ref{fig: fig1_cont}).
If it is located in the SNR, the accelerated particles can directly hit on it without escape, and thus hadronic emission from this clump is facilitated. 
Another reason may be that the escape of particles is non-isotropic due to the magnetic field configuration \citep[e.g.,][]{Nava2013}. In this case, Clump N could be located just in the escaping cone and illuminated by the energetic particles.

\section{Summary}\label{sec:conclusion}
We investigate the molecular environment of SNR \snr\ with FUGIN archival \twCO\ and \thCO\ data and our PMOD observation in \HCOp\ and HCN lines.
We also analysis GeV \gray\ emission possibly associated with the SNR using the {\sl Fermi}-LAT observational data. 
The principal results are summarized in following:

\begin{enumerate}
    \item SNR \snr\ is closely surrounded by a molecular belt in the southeastern boundary, with both SNR and belt recessed in the band-like molecular gas structure along the Galactic plane.
    
    \item 
    Asymmetric broad molecular line profiles (between $+10\km\ps$ -- $+30 \km\ps$) are extensively present in the molecular belt, both along the SNR boundary (regions R1, R2, R3) and in the outer regions (O1, O2), and the northern Clump N. The secondary components can be ascribed to the motion of the wind-blown shell. 
    This explanation is supported by the P-V diagram along a line across the remnant (between regions R2 and Clump N), which shows an arc-like pattern, suggesting an expanding gas structure. 
    
    \item  We obtain the abundance ratios $N$(\HCOp)/$N$(\twCO) for the five regions in the molecular belt and Clump N in the range $\sim 2.6\E{-5}$ -- $3.6\E{-4}$ on the LTE assumption.
    We simulate the chemical effects of shock propagation (including molecular ionization by penetrating LECR protons) with the Paris-Durham shock model, and show that
    the scenario of the shock of the wind-driven shell can more naturally account for \HCOp-to-\twCO\ abundance ratios than the scenario of SNR shock.

    \item  We suggest the southeastern molecular belt and northern clump are parts of an incomplete molecular shell of bubble, which was driven by O-type progenitor star's wind and is interacting with the SNR. 
    
    \item Based on the {\sl Fermi}-LAT 16-yr data, we find a possible \gray\ point source (`Source A') at R.A.$=275.9203^\circ$, Dec.$=-12.3349^\circ$ with a significance about 6.5$\sigma$ in 0.2--500 GeV. The photon index of the emission is 2.35$\pm$ 0.11, and the luminosity in 0.2--500 GeV is 1.8$\E{34}$ erg $\ps$.
    The position of this source accords to the northern molecular Clump\,N, which could be responsible for hadronic interaction for generating the \gray\ emission. 
    Our spectral fit of the emission indicates that a hadronic origin is favored by the measured Galactic number ratio between CR electrons and protons $K_{\rm ep}\sim0.01$.
    
\end{enumerate}


\section*{Acknowledgements}
The authors thank Ping Zhou and Yi-Heng Chi for helpful comments.
This work is supported by the NSFC under grants 12173018, 12121003, and 12393852.


\section*{Data Availability}
This work makes use of data from FUGIN, FOREST Unbiased Galactic plane Imaging survey with the Nobeyama 45 m telescope, a legacy project in the Nobeyama 45 m radio telescope of CO. 
The Fermi-LAT data underlying this work are publicly available, and
can be downloaded from \url{https://fermi.gsfc.nasa.gov/ssc/
data/access/lat/}.



\bibliographystyle{mnras}
\bibliography{kes67}


\appendix
\section{The Age of SNR Kes\,67}\label{apd: new age}

Using a blast wave velocity estimated from ram pressure balance assumed for the shocked MCs, 
\citet{Dubner1,age} suggested a dynamical age $1\E{5}\yr$ for \snr.
When \xray\ emission from the SNR was detected later, an average temperature of the hot gas across the remnant was obtained as $kT_{\rm X}\sim0.4\keV$ \citep{Fe}. 
Thus we can derive the temperature of the postshock gas as
\(k\Ts\approx kT_{\rm X}/1.27\sim0.3\keV\)
for the Sedov case \citep{Sedov_solution}.
The shock velocity is therefore given by
\(\vs=\left[(16/3)k\Ts/(\bar{\mu}\mH)\right]^{1/2}
\sim510\km\ps\),
where $\mH$ is the hydrogen atom mass and $\bar{\mu}=0.61$ is the mean atomic weight. 
With a mean angular radius $7'$ adopted, the mean radius of the remnant is $\rs\sim29$\,pc for $d\sim14\kpc$.
Consequently, we obtain a Sedov age as 
$t = 0.4\rs/\vs\sim 22$\,kyr. 


\bsp	
\label{lastpage}

\end{document}